# Nanophotonic Optical Switching Using Digital Metamaterials and Photochromism

Apratim Majumder, Mark Mondol, Trisha L. Andrew and Rajesh Menon

*Abstract*—Optical switches are one of the most important elements of integrated photonics. Here, we designed, fabricated and characterized several nanophotonic optical switches (NOSs) in silicon that exhibit ultra-compact footprint, along with excellent extinction ratios and operating bandwidths. Simulations indicate that our best device of dimensions 3μm×3μm and 0.3 μm device height can provide about 10dB extinction ratio with a bandwidth of 12nm, centered at the design wavelength of 1550nm. Three-dimensional finite-difference time-domain analysis in conjunction with a modified version of direct binary search algorithm was used to design the nanophotonic device structure. The modulation is a result of optical modulation of the refractive index of a photochromic material and an optimized geometry of the digital metamaterials that comprise the device. The devices are CMOS fabrication compatible and examples of ones designed using multi-state geometric optimization that will lead to a new class of ultra-compact and multi-functional active integrated-silicon devices.

## I. Introduction

OPTICAL communication systems typically incorporate photonic-integrated circuit (PIC) elements fabricated on the silicon-on-insulator (SOI) platform due to a number of advantages such as low power consumption, high efficiency, small footprint and compatibility with complementary metal-oxide-semiconductor (CMOS) fabrication processes [1-2]. An optical modulator is one of the most critical components in such a system [3]. Most commonly, optical modulation in silicon has been realized in the past using the plasma dispersion effect, where a change in the real and imaginary parts of the refractive index of silicon is caused by a change in its concentration of free carriers [4]. However, this effect is fairly weak allowing for refractive index modulation of only ~ 0.002 [5,6]. In general, modulators are characterized by the extinction ratio per device length, defined as $10\log(I_{max}/I_{min})/L$ dB/μm, where $I_{max}$ and $I_{min}$ are the intensities transmitted in the ON and OFF states, respectively, and L is the length of the device. Other important characteristics include the operating bandwidth, insertion loss and switching speed. Monolithic modulators, using direct carrier injection in p-i-n junction devices were able to achieve 80% modulation over a 10μm device length [4], equivalent to 0.69 dB/μm [7]. Liu et. al. demonstrated modulation using carrier accumulation on either side of a dielectric layer inside a micrometer-sized waveguide. Liao et al. improved this work and a device capable of 20dB extinction ratio over a 3.45mm device length (corresponding to 5.8 ×$10^{-6}$dB/μm) was demonstrated. Although the device footprint was large, modulation speeds >1GHz were reported [8-10]. A smaller device using the same working principle with a 9dB extinction ratio achieved over 480μm device length, corresponding to 18.8 × $10^{-3}$dB/μm was later reported [10]. Modulation achieved using carrier depletion has also been reported [11-16], among which, Liao et. al. [12] reported the fastest data transmission at 40 Gbit/s with 1dB extinction ratio. However, since these devices typically employ Mach-Zehnder Interferometer (MZI) type phase shifters in their design, the device footprint is usually in the 1mm to 3.5 mm range, which corresponds to a range of 1 × $10^{-6}$dB/μm to 28.6 × $10^{-8}$dB/μm. Xu et. al. introduced active high-speed ring resonators [17-20] with extinction ratios as high as 15dB over device lengths of 12-15μm equivalent to about 1.25dB/μm. An electro-optic polymer-based modulator with 3dB modulation depth was recently demonstrated on a footprint of hundreds of microns, corresponding to ~3×$10^{-2}$dB/μm [21]. Recently, there have been advances in modulators utilizing two-dimensional materials such as graphene [22-24], transition metal dichalcogenides [25-27], black phosphorus, etc. Of these, graphene performs exceptionally well, due to its high carrier mobility achieving extinction ratio per active length as high as 3dB/μm [24]. Different types of all-optical modulators have been demonstrated, such as saturable absorbers [28], optical limiters [29], polarization controllers [30] with large extinction ratios up to 27dB. However, these are discrete millimeter-scale devices and their on-chip counterparts have not been realized yet. A single atom plasmonic switch using silver with a 9.2dB extinction ratio was recently demonstrated [31]. However, extreme fabrication constraints as well as the use of non-CMOS compatible metals like silver and platinum required to realize this device limits its application. Most recently, Zhao et. al. demonstrated high all-optical modulation (60dB) using low power control signal, but on a millimeter-scale device, corresponding to < 0.06dB/μm [32]. Finally, optical modulation using ring resonators [33] is a very popular method and there have been examples of devices which have used the concept of ring resonators in conjunction with the

A. Majumder, R. Menon is with the Department of Electrical and Computer Engineering, University of Utah, Salt Lake City, UT 84102 USA (e-mail: apratim.majumder@ utah.edu; rmenon@eng.utah.edu).
M. Mondol, is with the Research Laboratory of Electronics, Massachusetts Institute of Technology, Cambridge MA 02139, USA (e-mail: mondol@mit.edu).
T. L. Andrew is with the Department of Chemistry, University of Massachusetts Amherst, Amherst MA 01003, USA (e-mail: tandrew@umass.edu).

electro-optic effect in Si [34] as well as the thermo-optic effect [35] to produce optical modulation. In these types of devices, high extinction ratio can be usually achieved using large ring radius, although the smallest device demonstrated is of comparable size to our device (radius ~ 1.5 µm) [33] and comparable performance, but not demonstrated to be tunable.

In contrast to previous work, here we designed, fabricated and characterized an all-optical switch that uses an optimized digital metamaterial geometry in silicon and a CMOS-compatible photochromic molecule that achieves modulation of 3.3dB/µm in an area of only 3µm×3µm with an operating bandwidth of about 12nm. The device may be fabricated in a single lithography step and can be fully CMOS compatible. Use of polymers to modulate optical signals is not new [36, 37]. However, the photochromic species used here has no absorption in the working wavelength range of the device. The modulation is achieved via an optimized arrangement of silicon and photochromic pixels, as determined by our inverse design algorithm. Indeed, the arrangement of silicon pillars alone, can also not achieve modulation, neither a random arrangement of the Si pillars with BTE filling the empty locations. Hence, we present an entirely new type of optical switching, one based on the principles of digital metamaterials, thereby allowing the device to maintain a small footprint. Another important point of this work is the first demonstration of using inverse design to optimize a multi-state (2-state here) device.

The mechanism for optical modulation is based, along with the specific pixel layout, upon the change in refractive index of a photochromic molecule by exposure to two different wavelengths of light [38-42]. Photochromes are low-density, soft molecular materials with widely adaptable processing requirements and can be introduced into a device stack at any point of a process flow, by spin-coating or through vapor state deposition. Films of the diarylethene photochromes used herein can be created using solvent-free physical vapor deposition, rendering this material CMOS compatible. A large extinction ratio per unit length is achieved due to the complex superposition of resonant guided modes that occur within our designed device as described below. Computationally designed nanophotonic devices have previously enabled a number of passive photonic functionalities [43-48] including free-space-to-waveguide couplers [43], optical diodes [45], polarization beamsplitters [46], densely packed waveguides [47], polarization rotators [48], etc. Here, we extend this computational method to active multi-state photonic devices using the example of an optical switch.

## II. Design of nanophotonic optical switches (NOS)

The device region (3µm × 3µm) is first divided into 30 × 30 square pixels, each of size 100nm × 100nm. Each pixel may be comprised of either silicon or of the photochromic material. Light of polarization state TE (electric field polarized in the plane of the device) is coupled into and out of the device via single-mode waveguides (see Fig. 1). The photochromic material can be switched from a closed-ring state (n=1.62 at λ=1.55µm) to an open-ring state (n=1.59 at λ=1.55µm), and vice-versa by illuminating it to red and near-UV wavelengths, respectively. This allows us to exploit a change in refractive index of 0.03 in the photochromic pixels, which is at least an order of magnitude higher than what is achievable in silicon using the plasma-dispersion effect [3]. We first arbitrarily choose the closed-ring isomeric state of the photochromic molecule as the ON state and the open-ring isomeric state as the OFF state of the modulator. This choice is not important as illustrated by a design using the opposite choice in Fig. S4 [52]. Details of the photochromic mechanism are further described elsewhere [52]. A nonlinear optimization is used to design the distribution of the silicon/photochromic pixels with the goal of maximizing a figure of merit (FOM), which in this case was the extinction ratio. We used a modified version of the direct-binary-search algorithm to implement the optimization (see section 2 of ref [52]).

The optimized geometry as well as cross sections through the device and through the waveguides are shown in Fig. 1. When the switch is illuminated from above with near-UV light (λ~325 nm), the photochromic pixels switch from the open-ring to the closed-ring state, and the switch enables transmission of the light from the input to the output waveguides. On the other hand, when the switch is illuminated with red light (λ~615 nm), the photochromic pixels undergo the opposite transformation, the modified refractive-index distribution results in low transmission of light from the input to the output.

In each state, coupled resonant modes are excited by the incident field, which either results in relatively high or low transmission to the output waveguide depending upon the refractive-index distribution (see Supplementary Movie 1). The steady state intensity distributions in the two states of the device are shown in Fig. 2 for λ=1550nm. Unlike traditional modulators, where light is manipulated via symmetry breaking of periodic structures, our devices use free-form metamaterials, where the distribution of the silicon and photochromic pixels is determined using optimization. From the simulations, it is revealed that for the OFF state of the device, most of the light is scattered out of plane and onto the sides of the device, while decreasing the amount of light that is coupled into the output waveguide. For the ON state of the device, coupled guided-resonant modes are generated that result in funneling of light to the output waveguide.

Since our design methodology relies on optimization, by emphasizing different terms in the figure-of-merit, it is possible to achieve devices with different characteristics. In order to emphasize this and the generality of our design approach, we present 3 other designs in ref [52]. Device 2 is shown in Fig. S2, and has -10.36dB insertion loss in the ON state compared to -15.3dB for the device in Fig. 1 (but at the expense of a small decrease in extinction ratio). Additionally, we present device 3 in Fig. S3, where we choose the starting design to have 65% Si pillars to ease the fabrication complexity. Again, we see that an optimized design is possible where the device behavior is similar to the first two designs. Finally, in Fig. S4, we present the design of device 4 where the ON and OFF states are reversed with respect to the refractive index of the photochromic material. Here, the open-ring state serves as ON, whereas the closed-ring state serves as OFF. We can see that an optimized design can be achieved. It is to be noted that our method is flexible enough to accommodate not only modifications to the figure-of-merit, but also more sophisticated algorithms such as the adjoint-matrix method or machine learning techniques.

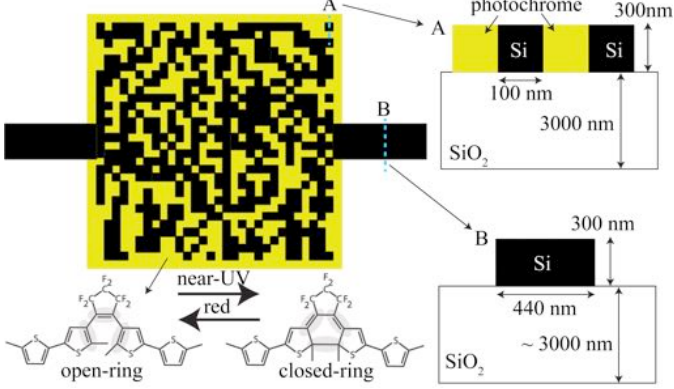

Fig. 1. The nanophotonic-optical switch. The optimized geometry of the switch is shown (black is silicon and photochrome is shown in yellow). Cross-sections A and B show the metamaterial pixel and the single-mode-waveguide details, respectively. The two isomeric forms of the photochrome are shown on the bottom left and are accessible by exposure to near-UV and red wavelengths as indicated.

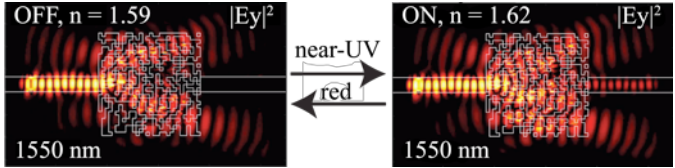

Fig. 2. Simulation of the nanophotonic-optical switch. Simulated steady-state intensity distribution of the OFF and ON states of the switch for $\lambda=1550$nm and TE polarization are shown. As explained in the text, the device is switched between these states by modulating the photochromic pixels via exposure to near-UV and red wavelengths of light.

## III. EXPERIMENT AND DEVICE CHARACTERIZATION

Each device was fabricated in an SOI wafer with silicon thickness of 300nm and oxide thickness of about 3000nm (see Fig. 1). The SOI wafer was prepared as described in section 3 of ref [52]. Due to our process constraints, we used polysilicon for the device layer, resulting in higher than normal scattering losses [50]. As mentioned earlier, the device may be fabricated in a single lithography step, since the height of the pixels and that of the waveguides is the same. We patterned 100nm thick hydrogen silsequioxane (HSQ) resist via scanning-electron-beam lithography (Elionix ELS-F125 125 keV) and developed with 1% Sodium Hydroxide, 4% Sodium Chloride in aqueous solution. The resulting pattern was etched into silicon by RIE (Plasma Therm) with Cl at 20 mTorr. Subsequently, the photochromic material was spin coated directly onto the device. On-chip polarizers were included to ensure that the input polarization state was transverse electric (TE) as assumed during design. Reference devices (waveguide with no switch) were also included on the same wafer for reference measurements. The scanning-electron micrograph of a fabricated device prior to the photochromic overcoating is shown in Fig. 3a. Finally, the devices were characterized using a setup that we have described previously [43-48] (also included in section 4 of ref [52]). Switching between the ON and OFF states was achieved by illuminating the entire chip with red and near-UV light from above. The transmission efficiency was computed relative to the reference device to account for any inadvertent polarization rotation that occurs during the free-space-to-waveguide coupling and for any losses in the taper from multi-mode to single-mode waveguides. Micrographs of the devices overcoated with the photochromic material are shown in Fig. S5. Devices described earlier as designs 2 and 3 were also fabricated and their details are included in Fig S6.

We plotted the relative transmission-efficiency spectrum, which is computed as the ratio of power in the output waveguide to that in a reference waveguide (without the switch), in Fig. 3b. The extinction ratio is defined as the ratio of the output power in the ON state to that in the OFF state for the same input power, and is plotted in Fig. 3c as a function of wavelength. The simulated and measured peak-extinction ratios are 10dB at $\lambda = 1.55\mu m$ and 8dB at $\lambda = 1.56\mu m$, respectively. If we define the operating spectral bandwidth as the wavelength range over which the extinction ratio is maintained above 30% of its peak value, then we estimate the simulated and measured bandwidths as 12nm and 10nm, respectively. The experimental data is shown with dashed lines, while the sold lines show the simulated plots. It is interesting to note that the polarity of the switch changes at $\lambda=1.58\mu m$, where absolute extinction ratio over 9dB is observed.

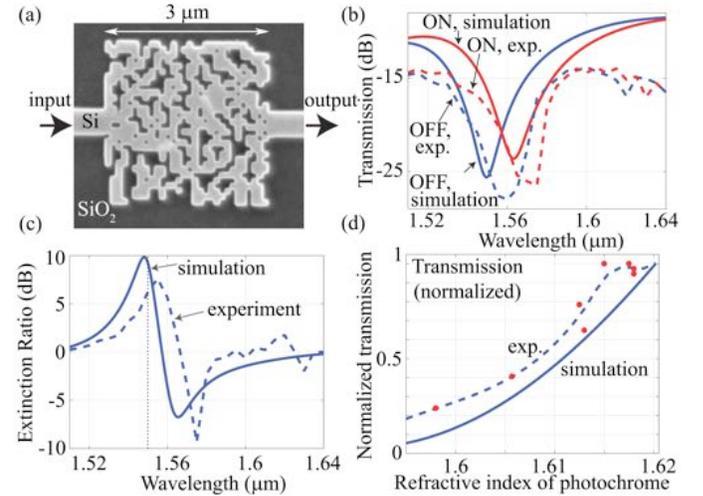

Fig. 3. Experimental results. (a) Scanning-electron micrograph of fabricated device before coating with the photochromic layer. (b) Measured and simulated transmission spectra for the ON and OFF states of the device. (c) Measured and simulated extinction-ratio spectra. (d) Measured (red filled circles are measurements, while blue dashed line is fit line) and simulated transmission (normalized to the peak value) as a function of the refractive index of the photochrome, showing the analog operation of the switch. In all curves, dashed lines denote experiments, while solid lines show simulations.

The refractive index of the photochrome can be changed continuously from 1.62 to 1.59 (at $\lambda=1.550\mu m$), and vice-versa by exposing it to visible and near-UV wavelengths. This allows us to continuously change the transmission of the switch, *i.e.*, turning it into an analog modulator. The simulated and measured (normalized to peak value) transmission as a function of the refractive index of the photochrome is shown in Fig. 3d. In order to experimentally characterize this, we first spin-coated a thin uniform layer of the photochrome on a bare silicon wafer. Then, this film was exposed to red light ($\lambda\sim650$nm, PAR38 26W LED lamp) until the photostationary state is reached. The refractive index of this layer was measured and recorded using a spectroscopic ellipsometer (Woollam). Then, the sample was exposed to a UV light source (λ~325nm, UVP UVGL-25 4W 0.16A UV Lamp) and the refractive index was monitored as a function of time, resulting in the plot of refractive index vs exposure dose (Fig. S1f). Next, the transmission of the switch was also measured as a function of the exposure dose in the same manner and thereby, we could obtain the transmission of each switch as a function of the refractive index of the photochrome. It is noted that the measured data qualitatively follows the simulations. Therefore, our device is also able to function as an analog modulator. There is good qualitative agreement between experiments and simulations for all 3 fabricated switches (see Fig. S6 for devices 2 and 3).

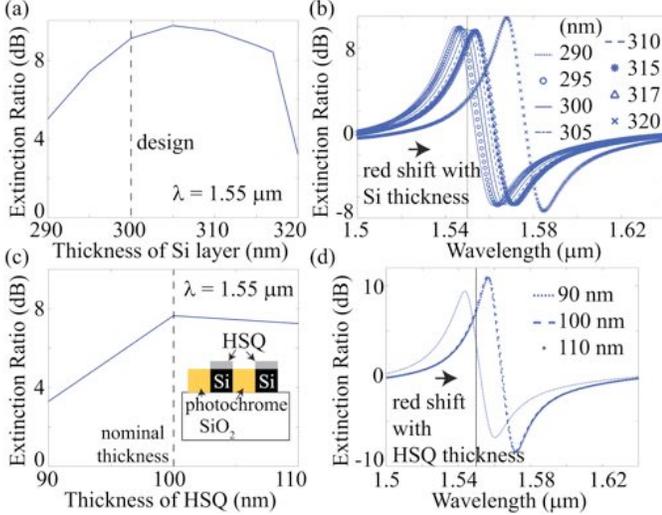

Fig. 4. Simulated effect of layer thicknesses. (a) Extinction ratio at λ=1.55μm as a function of the thickness of Si (device) layer and (b) the corresponding extinction-ratio spectra. (c) Extinction ratio at λ=1.55μm as a function of the thickness of top HSQ layer (as illustrated by the inset) and (d) the corresponding extinction-ratio spectra.

In order to understand the differences between experiments and simulations, we performed a numerical analysis of the impact of the silicon layer. As the silicon layer becomes thinner, the extinction ratio at λ=1.55μm decreases (Fig. 4a) as expected because less optical power is confined in the silicon pixels. However, the extinction ratio at λ=1.55μm increases a little with increasing Si thicknesses before decreasing as expected. Note that the design is optimized for Si thickness of 300nm. The extinction-ratio spectra in Fig. 4b shows the red-shift as Si thickness increases. To simplify our fabrication process, we did not remove the HSQ resist (nominal thickness of 100nm) from atop the silicon pixels as illustrated in the inset of Fig. 4c. We simulated the impact of this residual HSQ top-layer by computing the extinction ratio at λ=1.55μm as a function of the HSQ thickness. As indicated in Fig. 4d, the extinction ratio spectra are red-shifted for increasing HSQ thicknesses. We hypothesize that these two effects explain the observed red shift of the extinction-ratio spectra in our device. The reduction in peak extinction ratio can be attributed to small fabrication errors in the device geometry itself.

## IV. ANALYSIS OF DEVICE

It can be seen from Fig. 3(b) and 3(c) that the behavior of our NOS resembles that of a ring resonator with low Q-factor. We did not impose any restrictions on the behavior when implementing the optimization and it seems that the optimization converged on a traditionally known behavior. Analyzing the transmission and extinction characteristics, thus gives us an insight into the working of the device. It is evident that the arrangement of the Si pillars along with the photochromic material creates resonant modes inside the device. This creates the characteristic resonant transmission profile which matches closely with what is observed for ring resonators. When the refractive index of the photochromic device is modulated, there is a shift in the resonance of the device and this creates the shift in the minima of the transmission curves, thereby giving us the desired modulation. It is very interesting to note that a freely optimizing algorithm with no limiting conditions converged on a well-understood physical solution for our device. There is an inherent loss in our devices which is hard to limit given the multiple-holey nature of the device that gives rise to leaky modes. We have simulated different random geometries and have seen that an inherent loss of at least 5dB is always present in the present layout of the digital metamaterial (DMM). Finally, one may argue that a ring resonator with the photochromic material as the modulation medium may yield similar results. We show in Fig. 5 that a ring resonator modulated in such a way, with dimensions similar to our device, shows similar performance trend, albeit at a much worse extinction ratio. In other words, our optimized digital metamaterials device performs better than just a simple ring resonator.

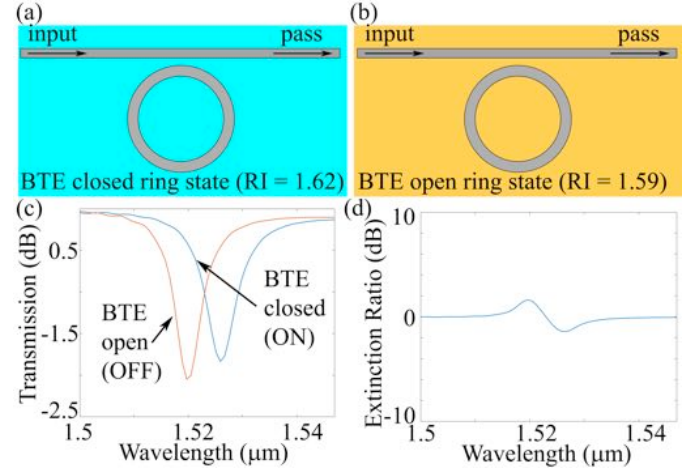

Fig. 5. Schematic of hypothetical ring resonator with the photochromic material as the refractive index modulating medium in (a) closed ring state and (b) open ring state. (c) Simulated transmission and (d) Extinction ratio of this device. The radius of the ring resonator is 1.5 μm and gap between the ring and the bus waveguide is 340nm.

## V. CONCLUSION

In summary, we designed, fabricated and characterized nanophotonic switches enabled by computationally designed digital metamaterials. We use an organic photochrome to optically modulate the refractive index, resulting in the switching behavior. The switch is only 3μm × 3μm in size and can provide extinction ratio per unit length of 3.3dB/μm over a bandwidth of 12nm. Perhaps most importantly, these devices are CMOS compatible and can be patterned in a single lithography step. Nevertheless, we emphasize that our design methodology can be applied to any mechanism of refractive-index modulation, thereby enabling a general principle for the design of ultra-compact switchable integrated photonics devices.

ACKNOWLEDGMENT

We thank Brian Baker, Tony Olsen and Steve Pritchett at the Utah Nanofab for assistance with sample preparation.

REFERENCES

[1] Almeida, V.R. et al. All-optical control of light on a silicon chip. Nature **431**, 1081-1083 (2004).
[2] Nagarajan, R. et al. Large-scale photonic integrated circuits. IEEE J. Sel. Top. Quantum Electron. **11**, 50–65 (2005).
[3] Reed, G.T. et al. Silicon optical modulators. Nat. Photonics **4**, 518-526 (2010).
[4] Reed, G. T. and Knights, A. P. Silicon Photonics: An Introduction Ch. **4**, 97–103 (Wiley, 2004).
[5] Reed, G. T. Silicon Photonics: The State of the Art Ch. 4, 95–145 (Wiley, 2008).
[6] Soref, R. and Bennett, B. Electrooptical effects in silicon. IEEE J. Quant. Electron. **23**, 123–129 (1987).
[7] Barrios, C. A., Almeida, V. R., Panepucci, R. & Lipson, M. Electrooptic modulation of silicon-on-insulator submicrometer-size waveguide devices. J. Lightwave Technol. 21, 2332–2339 (2003).
[8] Liu, A. et al. A high-speed silicon optical modulator based on a metal–oxide–semiconductor capacitor. Nature **427**, 615–618 (2004).
[9] Liao, L. et al. 40 Gbit/s silicon optical modulator for high-speed applications. Electron. Lett. **43**, 1196–1197 (2007).
[10] F. Y. Gardes, D. J. Thomson, N. G. Emerson and G. T. Reed, "40 Gb/s silicon photonics modulator for TE and TM polarisations," Opt. Exp. 19 (12) 11804-11814 (2011).
[11] Liu, A. et al. High-speed optical modulation based on carrier depletion in a silicon waveguide. Opt. Express **15**, 660–668 (2007).
[12] Liao, L. et al. 40 Gbit/s silicon optical modulator for high-speed applications. Electron. Lett. 43, 1196–1197 (2007).
[13] Liao, L. et al. High speed silicon Mach-Zehnder modulator. Opt. Express **13**, 3129–3135 (2005).
[14] Thomson, D. J. et. al. High contrast 40Gbit/s optical modulation in silicon. Opt. Express **19** (12), 11507–11516 (2011).
[15] Gardes, F. Y. et. al. A sub-micron depletion-type photonic modulator in silicon on insulator. Opt. Express **13**, 8845–8854 (2005).
[16] Green, W. M. et. al. Ultra-compact, low RF power, 10 Gb/s silicon Mach–Zehnder modulator. Opt. Express **15**, 17106–17113 (2007).
[17] Xu, Q. et. al. 12.5 Gbit/s carrier-injection-based silicon micro-ring silicon modulators. Opt. Express **15**, 430–436 (2007).
[18] Xu, Q. et. al. Micrometre-scale silicon electrooptic modulator. Nature **435**, 325–327 (2005).
[19] Manipatruni, S. et. al. High speed carrier injection 18 Gb/s silicon micro-ring electro-optic modulator. IEEE Proc. Lasers and Electro-Optics Soc. 537–538 (2007).
[20] Dong, P. et al. Low Vpp, ultralow-energy, compact, high-speed silicon electrooptic modulator. Opt. Express **17**, 22484–22490 (2009).
[21] Zhang, X. et. al. High performance optical modulator based on electro-optic polymer filled silicon slot photonic crystal waveguide. J. Lightwave Tech. **34** (12), 2941-2951 (2016).
[22] Liu, M. et al. A graphene-based broadband optical modulator. Nature **474**, 64–67 (2011).
[23] Gao, Y. et al. High-speed electro-optic modulator integrated with graphene-boron nitride heterostructure and photonic crystal nanocavity. Nano Lett. **15**, 2001–2005 (2015).
[24] Phare, C. T et al. Graphene electrooptic modulator with 30 GHz bandwidth. Nature Photon. **9**, 511–514 (2015).
[25] Gholipour, B. et. al. An all-optical, non-volatile, bidirectional, phase-change meta-switch. Adv. Mater. **25**, 3050–3054 (2013).
[26] Tittl, A. et. al. A switchable mid-infrared plasmonic perfect absorber with multispectral thermal imaging capability. Adv. Mater. **27**, 4597–4603 (2015).
[27] Karvounis, A. et. al. All-dielectric phase-change reconfigurable metasurface. Appl. Phys. Lett. **109**, 051103 (2016).
[28] Hasan, T. et al. Nanotube-polymer composites for ultrafast photonics. Adv. Mater. **21**, 3874–3899 (2009).
[29] Wang, J. et. al. Broadband nonlinear optical response of graphene dispersions. Adv. Mater. **21**, 2430–2435 (2009).
[30] Bao, Q. et al. Broadband graphene polarizer. Nature Photon. **5**, 411–415 (2011).
[31] Emboras, A. et. al. Atomic scale plasmonic switch. Nano Lett. **16**, 709-714 (2016).
[32] Zhao, H. Metawaveguide for asymmetric interferometric light-light switching. Phys. Rev. Lett. **117**, 193901 (2016).
[33] Q. Xu, D. Fattal, and R. G. Beausoleil, "Silicon microring resonators with 1.5-μm radius," Opt. Exp. 16 (6), 4309-4315 (2008).
[34] Q. Xu, B. Schmidt, S. Pradhan, and M. Lipson, "Micrometre-scale silicon electro-optic modulator," Nature 435 (7040), 325-327 (2005).
[35] M. S. Nawrocka, T. Liu, X. Wang, and R. R. Panepucci, "Tunable silicon microring resonator with wide free spectral range," App. Phys. Lett. 89 (7) 071110 (2006).
[36] Yacoubian, A. and Aye, T. M. Enhanced optical modulation using azo-dye polymers. Appl. Opt. **32**, 17, 3073-3080 (1993).
[37] Hochberg, M. et. al. Terahertz all-optical modulation in a silicon–polymer hybrid system. Nature Materials **5**, 703–709 (2006).
[38] Bianco, A. et. al. New developments in photochromic materials showing large change in the refractive index. Proc. of SPIE Vol. 7051, 705107, (2008).
[39] Andrew, T. L. et. al. Confining light to deep subwavelength dimensions to enable optical nanopatterning. Science, **324**, 917-921 (2009).
[40] Majumder, A. et. al., "Barrier-free absorbance modulation for super-resolution optical lithography", Opt. Exp., **23** (9), 12244-12250 (2015).
[41] Majumder, A. et. al., A comprehensive simulation model of the performance of photochromic films in absorbance-modulation-optical-lithography, AIP Advances, **6**, 035210, 2016.
[42] Majumder, A. et. al., Reverse-absorbance-modulation-optical lithography for optical nanopatterning at low light levels, AIP Advances, **6**, 065312, 2016.
[43] Shen, B. et al. Integrated metamaterials for efficient and compact free-space-to-waveguide coupling. Opt. Express **22**, 27175–27182 (2014).
[44] Shen, B. et al. Metamaterial-waveguide bends with effective bend radius λ₀/2. Opt. Lett. **40** 5750–5753 (2015).
[45] Shen, B. et al. Integrated digital metamaterials enables ultra-compact optical diodes. Opt. Express **23**, 10847–10855 (2015).
[46] Shen, B. et al. An integrated-nanophotonics polarization beamsplitter with 2.4 × 2.4 μm² footprint. Nat. Photonics **9** 378–382 (2015).
[47] Shen, B. et al. Increasing the density of passive photonic-integrated circuits via nanophotonic cloaking. Nat. Commun. **7** 13126 (2016).
[48] Majumder, A. et al. Ultra-compact polarization rotation in integrated silicon photonics using digital metamaterials. Opt. Express **25**, 19721–19731 (2017).
[49] Irie, M. Diarylethenes for memories and switches. Chem. Rev. **100**, 1685–1716 (2000).
[50] Black, M.R. Losses in polysilicon waveguides. Thesis (M. Eng.) Massachusetts Institute of Technology, Dept. of Electrical Engineering and Computer Science (1995).
[51] S. Shim, T. Joo, S. C. Bae, K. S. Kim and E. Kim, Ring opening dynamics of a photochromic diarylethene derivative in solution, J. Phys. Chem. A. 107(40), 8106-8110 (2003).
[52] Supplementary information.

# Nanophotonic Optical Switching Using Digital Metamaterials and Photochromism

Apratim Majumder, Mark Mondol, Trisha L. Andrew and Rajesh Menon[1]

Supplementary Material

I. DETAILS OF PHOTOCHROMIC MATERIAL

We used a specific photochromic material as the switchable pixels of the nanophotonic all-optical switch (NOS). The photochrome used in this work belongs to the diarylethene family. Specifically, it is the 1,2-bis (5,5' –dimethyl-2,2'-bithiophen-4-yl) perfluorocyclopent-1-ene molecule (otherwise referred to as BTE). We have previously used this material in demonstrating super-resolution photolithography [1-4]. The chemical structure of the molecule is shown in Fig. S1a. The molecule has the ability to reversibly transition between two isomeric states with different molecular structures, based on the wavelength of the photon absorbed by each state. We call these two isomers, the "open" and the "closed" forms, based on the structure of the central benzene ring in the respective isomer. The open form converts to the closed form by absorbing photons around 320 nm wavelength and the reverse state transition is initiated by the closed form absorbing photons at 615 nm. We use out-of-plane illuminations at these two wavelengths to induce the state changes of the photochromic pixels. The UV-Vis spectrophotometric analysis of the molecule showing its absorbance spectra is presented in Fig. S1b. At the same time, these two isomeric forms possess different refractive indices (real part of refractive index, $n_{Open}$ for the open form and $n_{Closed}$ for the closed form) at the communication wavelength $\lambda = 1.55 \mu m$. This has been reported previously [5] and we have independently confirmed this observation by measuring the refractive indices of the two forms using ellipsometric analysis on a Woollam Variable Angle Spectroscopic Ellipsometer (VASE). This is shown in Fig. S1c. For the NOS, the two different values of the real part of the refractive index of the photochrome at $\lambda = 1.55 \mu m$ govern the two states ("ON" and "OFF") of the modulator. The difference of the two values of refractive index at $\lambda = 1.55 \mu m$ ($\Delta n = n_{Closed} - n_{Open} = 1.62 - 1.59$) is 0.0301 as shown in Fig. S1d. This is significantly larger than the refractive index difference available in silicon by electrical effects by at least an order of magnitude [6]. Additionally, the photochromic material has no absorbance in the range $\lambda = 1.5 \mu m$ to $1.6 \mu m$ as seen in the ellipsometry data presented in Fig. S1e, which indicates that the NOS does not function via any absorbance of the input signal. The photochromic pixels in the NOS are defined by first using electron beam lithography and plasma etching to create empty pixels by etching away and thereby removing silicon, followed by spin coating the photochrome to fill these empty pixels and thereby create the photochromic pixels. In order to spin coat this material, the photochromic molecules are first dissolved by 93.63 weight percent in a polymer matrix of a 30 mg/mL solution of polystyrene dissolved in toluene. The process of synthesizing the molecules have been described elsewhere [1].

[1] A. Majumder, R. Menon is with the Department of Electrical and Computer Engineering, University of Utah, Salt Lake City, UT 84102 USA (e-mail: apratim.majumder@ utah.edu; rmenon@eng.utah.edu).
M. Mondol, is with the Research Laboratory of Electronics, Massachusetts Institute of Technology, Cambridge MA 02139, USA (e-mail: mondol@mit.edu).
T. L. Andrew is with the Department of Chemistry, University of Massachusetts Amherst, Amherst MA 01003, USA (e-mail: tandrew@umass.edu).

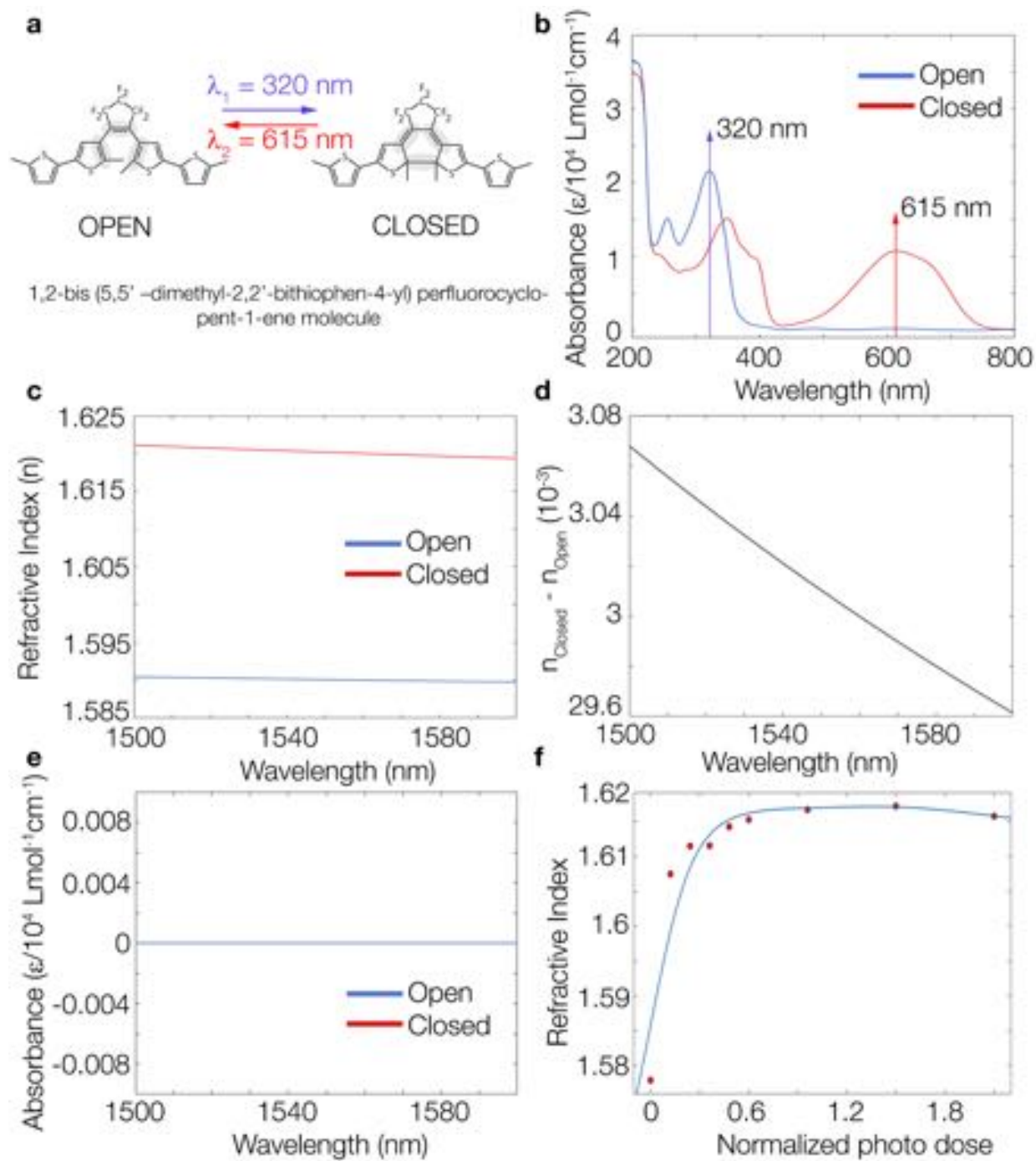

**Figure S1. Photochromic material (a)** Molecular structures of the open and closed isomers of the photochromic molecule showing the state change due to absorbance of UV and visible photons [1] **(b)** UV-Vis spectrophotometry data showing the absorbance of the photochromic material with peaks at 320 nm for the open form and 615 nm for the closed form **(c)** Ellipsometry data showing two different refractive indices for the two isomers of the molecule.

Independently measured and confirmed with [5] **(d)** Difference in the refractive index of the two forms **(e)** UV-Vis spectrophotometry data showing no absorbance of the photochromic material in the wavelength range 1500 to 1600 nm **(f)** Measured refractive index v/s normalized photo dose.

## II. Details of Design

Similar to our previous work [7-12], here, we used a modified version of the direct-binary-search (DBS) algorithm [13,14] to design the nanophotonic all optical modulator. One of the main advantages of using this algorithm over other similar yet distinct optimization techniques like generic algorithm and inverse design [15,16] is the fact that DBS has the ability to discretize the space over which the desired electromagnetic phenomenon takes place and thereby produce structures with discrete pixels, thereby facilitating fabrication in the nanoscale using conventional techniques like electron beam (e-beam) lithography, as we used here, or focused ion beam (FiB) milling, which we have used in the past [12]. Discrete pixels make our devices easier to fabricate, compared to continuous structures as produced by the result of inverse design algorithms [15]. Also, this makes our device fabrication compatible with CMOS industry standard projection lithography that can fabricate sub-50 nm structures. In DBS, a given region of space is discretized in to a number of pixels. In this case, the region of space was a 3 μm × 3 μm area, discretized into 30 × 30 square pixels. Each pixel thus has dimension of 100 nm × 100 nm which also defines the size of the smallest feature that needs to be fabricated for our device. This is well within the reach of prevalent nanofabrication techniques like the FiB and e-beam lithography. The DBS algorithm performs a nonlinear optimization to arrive at a specific pattern for the pixels that achieves the desired electromagnetic phenomenon, which in this case is intensity modulation of an input signal. One of the disadvantages of DBS is its high computation time. Hence, we chose to parallelize the process using elastic cloud computing on Amazon Web Services servers. The other drawback is that at times DBS tends to converge prematurely due to its sensitivity to the initial conditions. As a result, we repeated the optimization process with a number of random initial designs generated by Matlab. However, we noted that most starting points converged to designs with similar performance. Nevertheless, we have presented 4 different designs in this work.

For the NOS, there are two types of pixels. These are silicon and the photochromic material. The latter again may exist in one of its two states with different refractive indices as mentioned in the previous section. We define a figure of merit (FOM) as the average transmission efficiency of the input signal through the NOS device. A random starting pattern for the pixel layout was generated by Matlab and then passed into the DBS algorithm. The FOM was calculated for this pattern with the photochromic material in its two states, open and closed, as mentioned in the previous section, with two different values of refractive index in the two states, 1.59 and 1.62, respectively. The FOM when the photochromic material is in the open state is '$FOM_{Open}$' and when it is in the closed state is '$FOM_{Closed}$'. Without any loss in generality, we attributed the high transmission state of the switch to when the photochromic pixels are in the closed state. Hence, a final FOM is calculated as $FOM_{Final} = FOM_{Closed}/FOM_{Open}$. Next, the pixel is toggled to the other material state, i.e. if the pixel was silicon in the previous step, it is toggled to photochromic material and if it was the latter, then it is toggled to silicon. Again, $FOM_{Final}$ is calculated. If the new value of $FOM_{Final}$ is larger than the previous value, then the current pixel material state is retained, otherwise switched back. Then the next pixel is toggled and this iterative optimization of the process continues. One iteration ends when all 900 pixels have been toggled. The total optimization terminates when the final FOM does not improve beyond a predefined threshold (<0.5% in this case). We used finite-difference-time-domain (FDTD) method for simulating the electromagnetic fields in the devices and calculating the FOM, similar our previous work [7-11]. A restriction on the insertion loss of the device can also be introduced into the FOM, which we did for Design 2 and it can be seen that for that design, the insertion loss is less. The first design is presented in the main text while three other designs are presented in Fig. S2, S3 and S4.

**Figure S2. Nanophotonic optical switch (NOS) – Design 2**. (a) Schematic showing the geometry of the NOM device. (b) Simulated steady-state intensity distributions for ON and OFF states of the device at the design wavelength of 1.550 μm (see Supplementary Movie 2) [49]. The white arrows in (a, b) show the direction of propagation of light through the device. (c) Simulated relative transmission of the NOM as a function of wavelength. (d) Simulated extinction ratio as a function of wavelength, showing a peak of 6.9 dB at λ=1.555 μm and about 6.263 dB at λ = 1.550 μm.

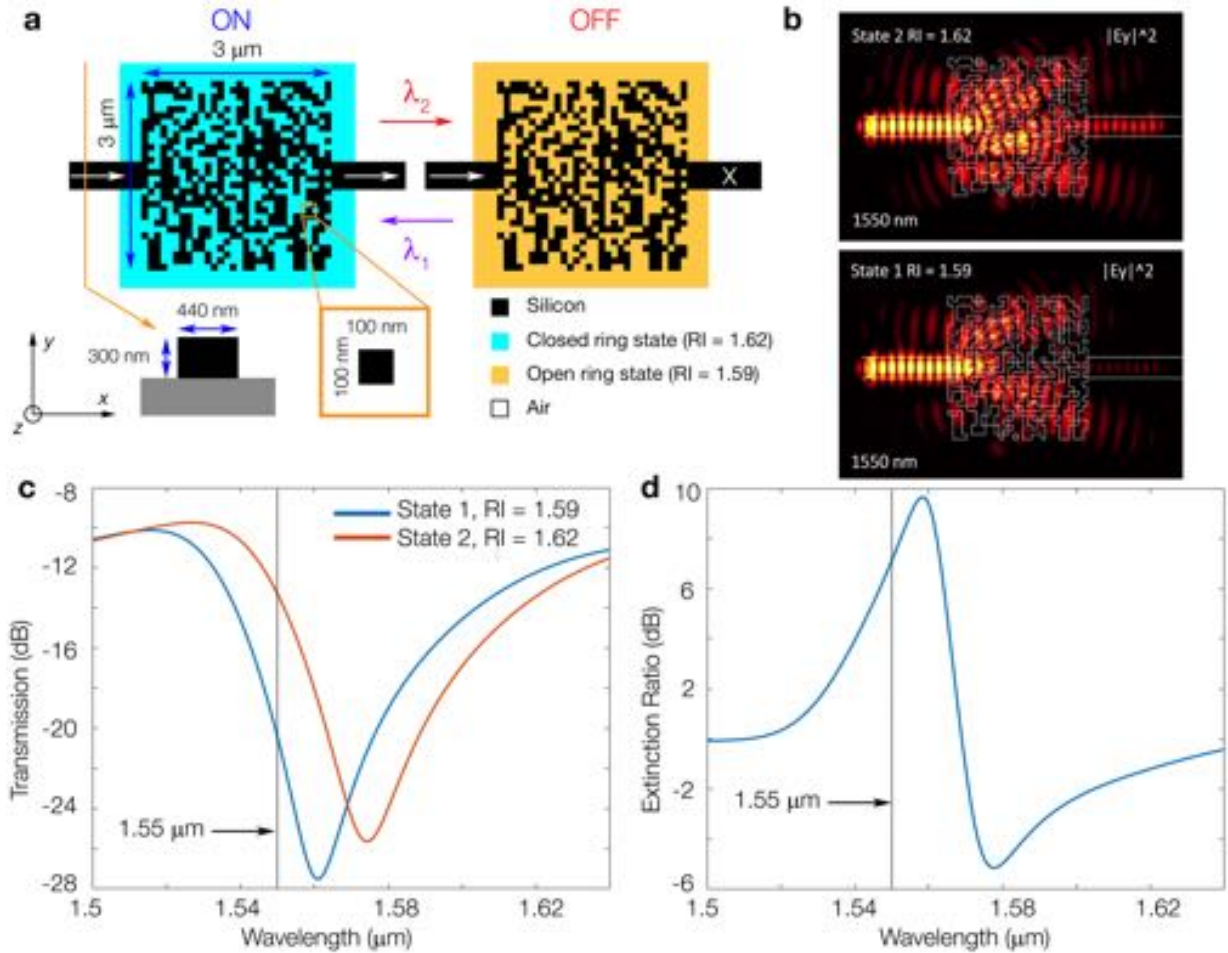

**Figure S3. Nanophotonic optical switch (NOS) – Design 3**. **(a)** Schematic showing the geometry of the NOM device. **(b)** Simulated steady-state intensity distributions for ON and OFF states of the device at the design wavelength of 1.550 μm (see Supplementary Movie 3) [49]. The white arrows in **(a, b)** show the direction of propagation of light through the device. **(c)** Simulated relative transmission of the NOM as a function of wavelength. **(d)** Simulated extinction ratio as a function of wavelength, showing a peak of 9.655 dB at λ=1.558 μm and about 7.206 dB at λ = 1.550 μm.

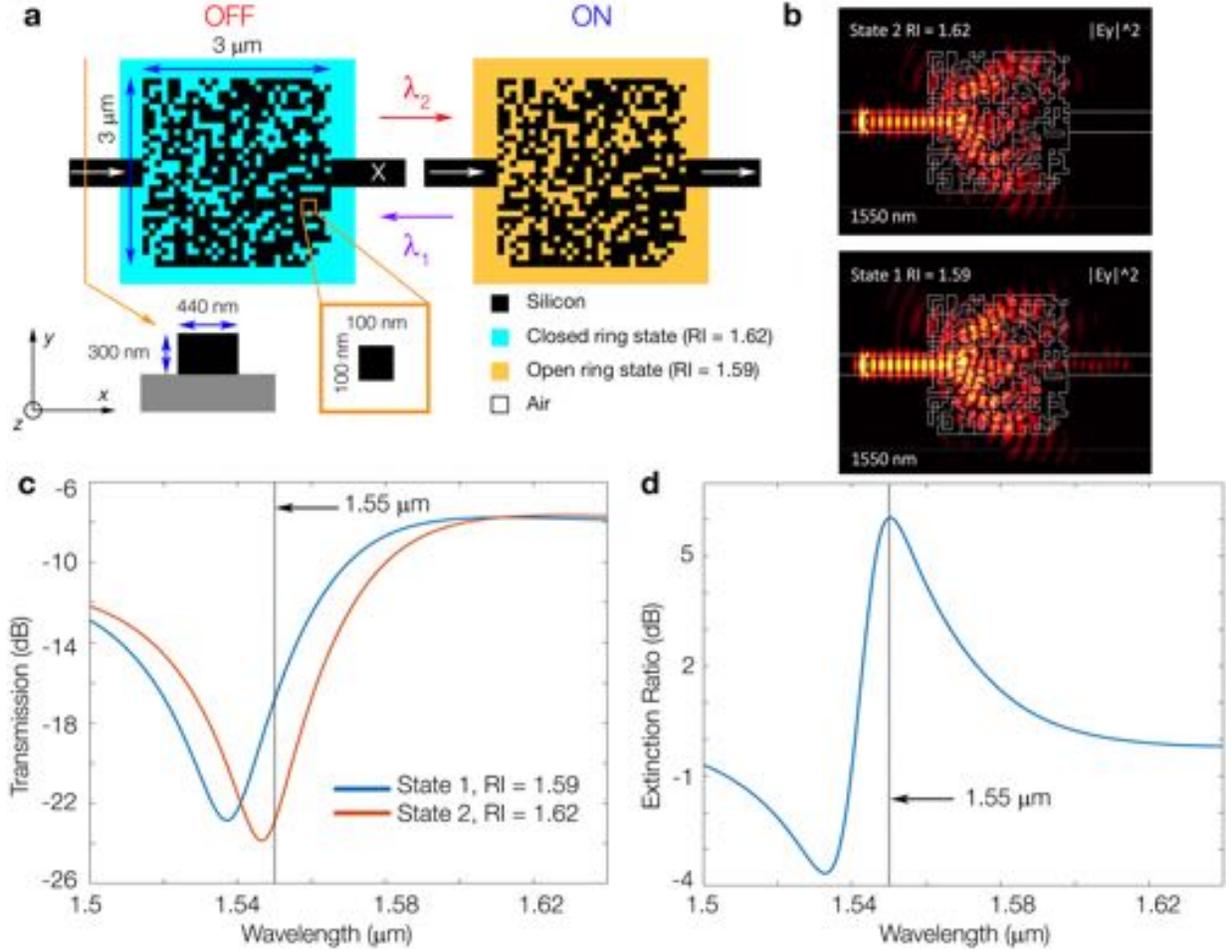

**Figure S4. Nanophotonic optical switch (NOS) – Design 4**. **(a)** Schematic showing the geometry of the NOM device. **(b)** Simulated steady-state intensity distributions for ON and OFF states of the device at the design wavelength of 1.550 μm (see Supplementary Movie 4) [49]. The white arrows in **(a, b)** show the direction of propagation of light through the device. **(c)** Simulated relative transmission of the NOM as a function of wavelength. **(d)** Simulated extinction ratio as a function of wavelength, showing a peak of 6.061 dB at λ=1.550 μm.

III. DETAILED SAMPLE PREPARATION METHOD

We fabricated the NOS device on the conventional platform of a silicon-on-insulator (SOI) wafer. As per our design constraint, the thickness of the top silicon layer was required to be 300 nm. However, since we were unable to procure SOI wafers with this exact silicon layer thickness at the time, we resorted to preparing our own wafer. In order to do this, first a 4-inch silicon wafer (University Wafers) was cleaned using acetone, methanol and isopropanol. Subsequently, the wafer was exposed to oxygen plasma on a Technics PE-II to remove all surface contaminants. Next, we deposited a ~ 3 μm thick silicon oxide ($SiO_2$) layer on the surface of this wafer using plasma enhanced chemical vapor deposition (PECVD) technique on an Oxford Plasmalab 80+. Lastly, undoped polycrystalline silicon (PolySi) of thickness ~ 300 nm was deposited in an Expertech furnace using the method of low pressure chemical vapor deposition (LPCVD), deposited at 630°C at a deposition rate of ~ 11.5 nm/min. The use of PolySi leads to undesired light scattering and contributes majorly to losses in the device. Since the height of the pixels in our device is the same as that of the input and output single mode waveguides, which is same as the height of the top silicon layer, our entire device with all waveguides and tapers may be fabricated in one single step. We patterned 100nm thick hydrogen silsequioxane (HSQ) resist via scanning-electron-beam lithography (Elionix ELS-F125 125 keV) and developed with 1% Sodium Hydroxide, 4% Sodium Chloride in aqueous solution. The resulting pattern was etched into silicon by RIE (Plasma Therm) with Cl at 20 mTorr. Reference devices for normalization including the same tapers but no device and on-chip polarizers [7-12] for setting the input polarization state were also fabricated by this same process on the same wafer. Next, the empty pixels defined up till this point needed to be filled with photochromic material. We resorted to spin coating the photochromic material onto the sample at this point, thereby filling the empty pixels. In order to do this, molecules of the BTE in the powder form were dissolved in a 30 mg/mL solution of polystyrene dissolved in toluene, by 93.63 weight percentage. This solution was prepared by stirring in an ultrasonicator for 4-5 hours. The sample with the waveguides and the NOS was first coated with a monolayer of HMDS to facilitate adhesion. Next, the photochromic solution was spin coated at 1000 rpm for 60 seconds with a prior 500 rpm spread for 5 seconds.

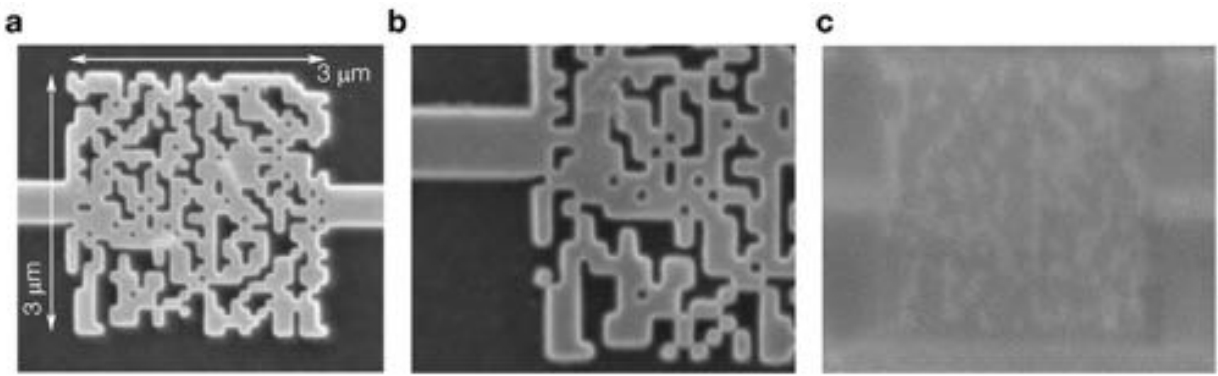

**Figure S5. Device fabrication results (a, b)** Scanning electron micrograph showing the fabricated NOS before and **(c)** after covering with photochromic material.

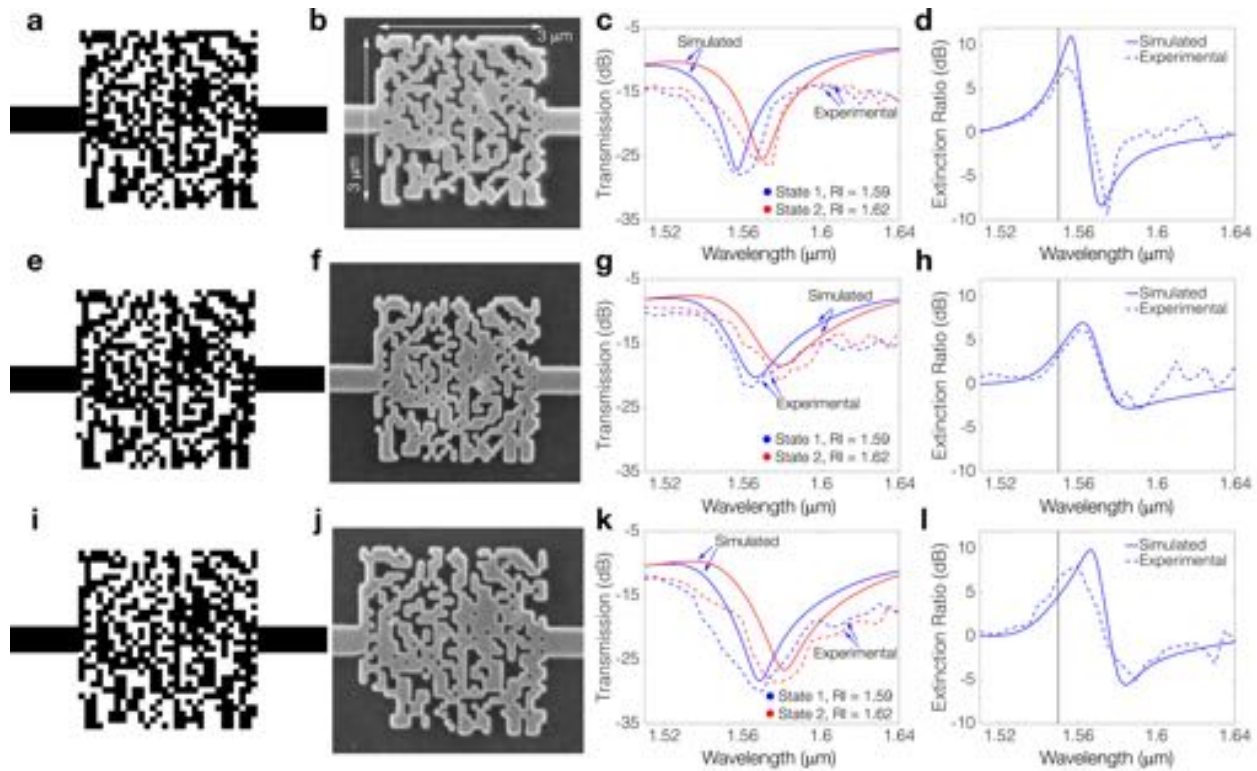

**Figure S6. Experimental results. (a-c)** Relative transmission and **(d-f)** Extinction ratio spectra are plotted for the devices 1, 2 and 3, respectively.

## IV. Device characterization

The setup used to characterize the fabricated NOS device is schematically shown in Fig. S7a. A standard single-mode lensed fiber was used at the input and a standard multi-mode fiber was used at the output to couple light into and from the multimode waveguides that then taper down to the single mode waveguides coupling in and out of the NOS device. The source was a Hewlett-Packard 8186F tunable laser source emitting in the range 1.508-1.640 μm. Similar to our previous work [7-11], we used an on-chip polarizer to set the polarization state of the input light. First, the single-mode input lensed fiber is aligned to the on-chip polarizer input. Butt-coupling is used to couple light from the lensed fiber to the waveguide that couples to the on-chip polarizer. The output is collected using a multi-mode lensed fiber. The fiber polarization controller PC1 is adjusted to set the input polarization state to TE by monitoring the power at the detector. The on-chip polarizer only allows the TM mode to pass through efficiently while blocking the TE mode. Hence, we could set the TE mode for the input by minimizing the power received at the detector.

Next, the input lensed fiber and output fiber could be moved to illuminate a straight reference waveguide with no device but all similar tapers or one with the NOS device in place. Once aligned, the measurements were recorded. For the NOS device, as shown in Fig. S7b, the device is first illuminated by a source with its emission centered at 320 nm ($\lambda \sim 325$nm, UVP UVGL-25 4W 0.16A UV Lamp) to convert the photochromic pixels to the closed state, with refractive index 1.62 and also high transmission for the switch. The output is collected. Next the device is illuminated by a source centered at 650 nm ($\lambda \sim 650$nm, PAR38 26W LED lamp) to convert the photochromic pixels to the open state, with refractive index 1.59 and thus, low transmission for the switch. The output is again collected. Thus, we could measure all the NOS devices fabricated.

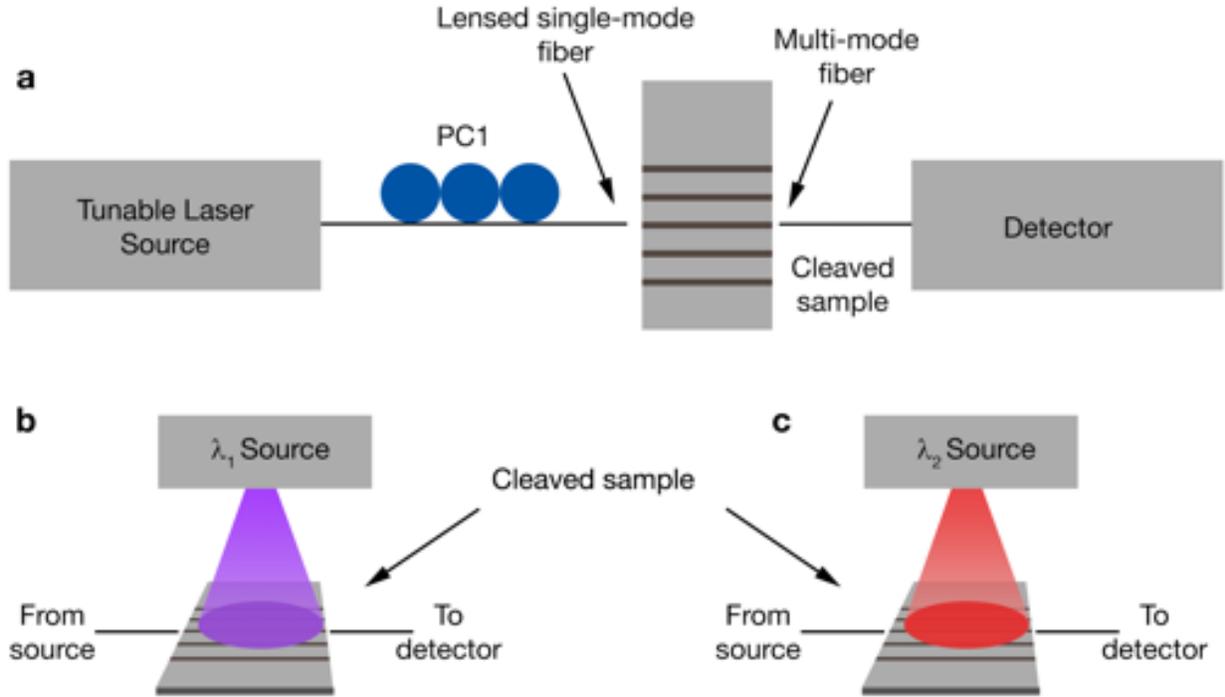

**Figure S7. Measurement setup** (a) Schematic of the measurement setup used to characterize the NOS device. The input signal is provided via a standard single-mode lensed fiber and the output is collected using a standard multi-mode fiber, in both cases using butt-coupling. (b) Schematic of the switching procedure showing the sample being illuminated by two sources emitting photons that can induce state changes in the photochromic material. Measurements are collected after each switching step.

V. ANIMATION DESCRIPTION

An animation is provided as supplementary information showing the simulated time evolution of the electric field pattern within the nanophotonic all optical switch design (shown in Fig. 1) for the on and the off states. The dimensions of the device are same as described in the manuscript. Light travels from left to right. The contour plot of the permittivity distribution of the device is shown using white lines.

**References**

1. T. L. Andrew, et. al., "Confining light to deep subwavelength dimensions to enable optical nanopatterning," Science, 324, 917-921 (2009).

2. A. Majumder, et. al., "Barrier-free absorbance switching for super-resolution optical lithography," Opt. Exp., 23 (9), 12244-12250 (2015).

3. A. Majumder, et. al., "A comprehensive simulation model of the performance of photochromic films in absorbance-switching-optical-lithography," AIP Advances, 6, 035210, 2016.

4. A. Majumder, et. al., "Reverse-absorbance-switching-optical lithography for optical nanopatterning at low light levels," AIP Advances, 6, 065312, 2016.

5. A. Bianco, et. al., "New developments in photochromic materials showing large change in the refractive index," Proc. of SPIE Vol. 7051, 705107, (2008).

6. G. T. Reed, et al., "Silicon optical switchs," Nat. Photonics 4, 518-526 (2010).

7. B. Shen, et al., "Integrated metamaterials for efficient and compact free-space-to-waveguide coupling," Opt. Express 22, 27175–27182 (2014).

8. B. Shen, et al., "Metamaterial-waveguide bends with effective bend radius λ0/2," Opt. Lett. 40 5750–5753 (2015).

9. B. Shen, et al., "Integrated digital metamaterials enable ultra-compact optical diodes," Opt. Express 23, 10847–10855 (2015).

10. B. Shen, et al., "An integrated-nanophotonics polarization beamsplitter with 2.4 × 2.4 µm2 footprint," Nat. Photonics 9 378–382 (2015).

11. B. Shen, et al., "Increasing the density of passive photonic-integrated circuits via nanophotonic cloaking," Nat. Commun. 7 13126 (2016).

12. A. Majumder, et al., "Ultra-compact polarization rotation in integrated silicon photonics using digital metamaterials," Opt. Express 25, 19721–19731 (2017).

13. G. Kim, et. al., "Increased photovoltaic power output via diffractive spectrum separation," Phys. Rev. Lett. 110 (12), 123901 (2013).

14. M. A. Seldowitz, et al., "Synthesis of digital holograms by direct binary search," Appl. Opt. 26 (14) 2788-2798 (1987).

15. A. Y. Piggott, et al., "Inverse design and demonstration of a compact and broadband on-chip wavelength demultiplexer," Nat. Photonics 9 374-377 (2015).

16. J. Lu, J. et al., "Nanophotonic computational design," Opt. Express 21 (11) 13351-13367 (2013).